\documentclass[amsmath,amssymb,notitlepage, twocolumn,aps,pra,floatfix]{revtex4-2}
\usepackage{xcolor}
\usepackage[colorlinks=true, allcolors = black]{hyperref}
\usepackage{url}
\usepackage{qcircuit}
\usepackage{subfigure}
\usepackage{graphicx}
\usepackage{dcolumn}
\usepackage{bm}
\usepackage{siunitx}

\usepackage{multirow}
\usepackage{braket}
\usepackage{amsmath}
\usepackage{cleveref}
\usepackage{mathtools}

\begin{document}
\preprint{APS/123-QED}

\title{Near-limit quantum control beyond analytic tractability in many-body spin systems\\}

\author{Jixing Zhang$^{1}$}
\thanks{These authors contributed equally to this work.}

\author{Bo Peng$^{2}$}
\thanks{These authors contributed equally to this work.}

\author{Yang Wang$^{1}$\textsuperscript{*}}
\thanks{ywang.qt@gmail.com}

\author{Cheuk Kit Cheung$^{1}$}
\thanks{These authors contributed equally to this work.}

\author{Guodong Bian$^{3}$}
\author{Hualuo Pang$^{2}$}
\author{Andrew M. Edmonds$^{4}$}
\author{Matthew Markham$^{4}$}
\author{Zhe Zhao$^{2}$}
\author{Yuan Hou$^{5}$}
\author{Durga Bhaktavatsala Rao Dasari$^{1}$}
\author{Ruoming Peng$^{1}$}
\author{Ye Wei$^{2,6}$}
\thanks{ye.wei@cityu.edu.hk}
\author{J\"{o}rg Wrachtrup$^{1,5}$}

\affiliation{$^{1}$3rd Institute of Physics, University of Stuttgart, Allmandring 13, Stuttgart, 70569, Germany}
\affiliation{$^{2}$Department of Data Science, City University of Hong Kong, Hong Kong, China}
\affiliation{$^{3}$School of Chemistry, University of Birmingham, B15 2TT, Edgbaston Birmingham, UK}
\affiliation{$^{3}$School of Chemistry, University of Birmingham, B15 2TT, Edgbaston Birmingham, UK}
\affiliation{$^{4}$Element Six Global Innovation Centre, Fermi Avenue, Harwell Oxford, Didcot, Oxfordshire OX11 0QR, United Kingdom}
\affiliation{$^{5}$ Max Planck Institute for Solid State Research, Heisenbergstraße 1, Stuttgart, 70569, Germany}
\affiliation{$^{6}$Department of Materials Science, City University of Hong Kong, Hong Kong, China}

\begin{abstract}
As quantum control approaches hardware-imposed performance limits, weak effects omitted by reduced models become consequential. Assumptions required for analytic tractability then cease to guide control design and instead constrain further improvement.
Here, we relax such assumptions and use simulation-guided stochastic tree search to navigate combinatorially large, discrete pulse-sequence spaces for robust many-body spin control. Experimentally, in a solid-state spin ensemble, the resulting computationally discovered pulse sequences substantially outperform analytically optimized baselines, despite being excluded by construction from analytic design criteria.
Importantly, these unconventional sequences expose predictive structural features that enable rapid neural network--based performance evaluation. This efficiency gain makes the combinatorial scaling tractable and expands the control alphabet from 8 symmetry-restricted pulses to over 26,000 hardware-resolved options. The resulting fine-grained design freedom provides the control resolution required to reliably address weak, performance-limiting effects, unlocking qualitatively different spin-control capabilities beyond decades of traditional sequence design.
Together, these results show that near performance limits, simplifying assumptions can become a primary constraint on quantum control in realistic hardware, and must be repurposed to guide computational discovery.
\end{abstract}

\maketitle
\section{Introduction}
As quantum technologies transition from proof-of-principle demonstrations toward engineered platforms, control performance increasingly enters regimes in which previously negligible effects become consequential~\cite{Haroche2006,Awschalom2025,Acharya2024,Bluvstein2025}. Assumptions that enable closed-form reasoning become progressively misaligned with weak residual effects that are omitted in tractable models yet dominate near-limit performance~\cite{Preskill2018,Brif2010}. This structural tension motivates computational methods capable of relaxing assumptions required for analytic tractability, allowing them to function as soft guidance rather than hard design constraints.

Computational approaches have enabled substantial progress across quantum technologies~\cite{Brif2010,Rabitz2000}, particularly for tasks with well-defined objectives and structured, learnable solution spaces~\cite{Bausch2024,Zhou2025a,Foesel2018,Gustafson2025}. However, certain central problems do not naturally meet these conditions and continue to resist systematic computational treatment, instead predominantly relying on closed-form analysis, notably periodic pulse sequences that engineer time-averaged quantum dynamics~\cite{Dyte2025,Choi2020PRX,Peng2022}.
This resistance reflects an underlying circularity: assumptions introduced to ensure analytic tractability also define available search strategies, thereby preventing them from relaxing these very assumptions~\cite{Simon1996,Brif2010,Merchant2023}.

\begin{figure*}[tbp]
\includegraphics[width=0.8\textwidth]{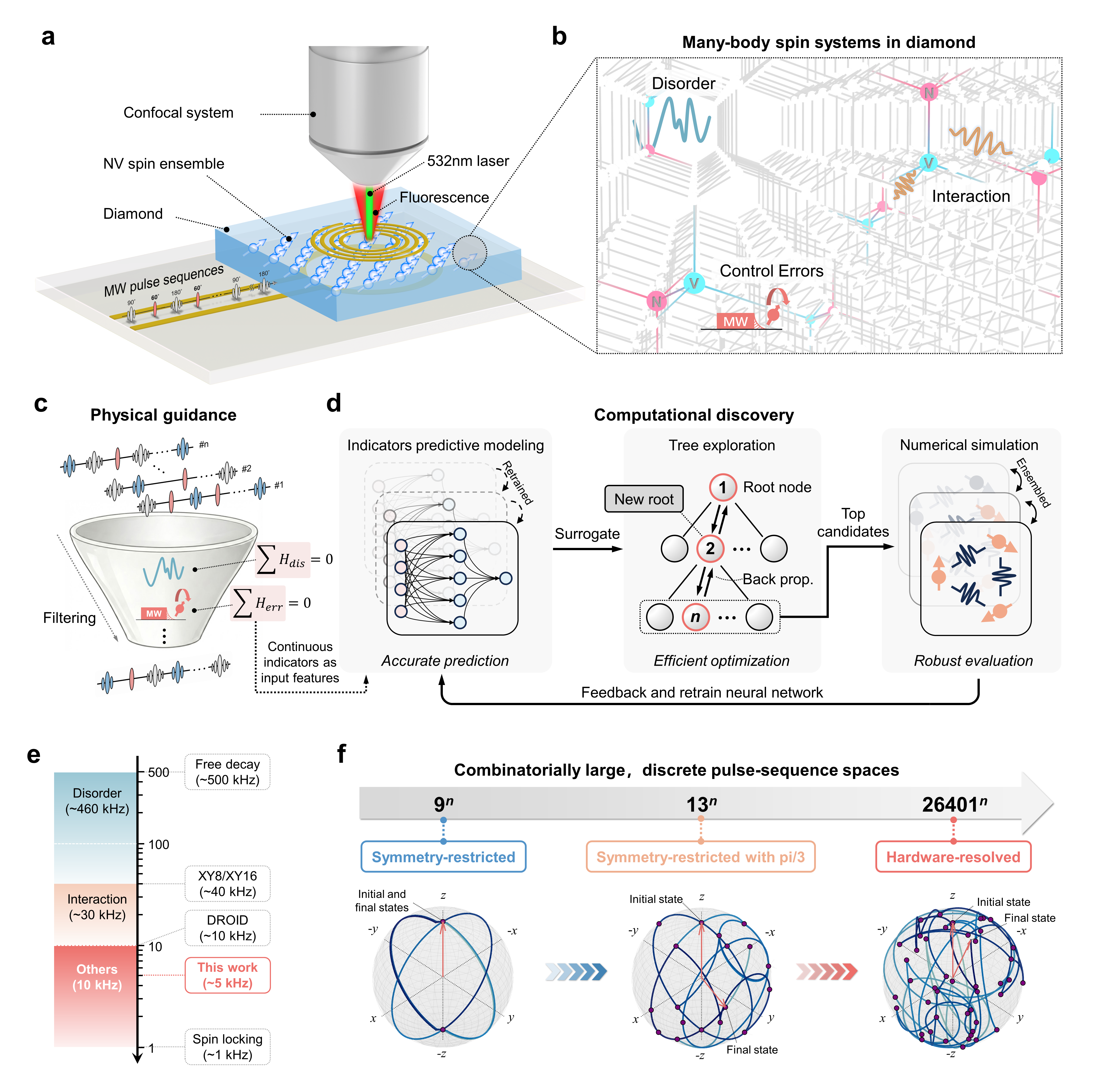}
\caption{
\textbf{Overview of physics-guided computational discovery for sequence design}.
\textbf{(a)} A dense diamond spin ensemble integrated with a room-temperature confocal optical system, enabling microwave (MW) pulse control and fluorescence-based readout.
\textbf{(b)} Disorder, interactions, and control imperfections give rise to the complex many-body spin dynamics.
\textbf{(c)} Traditional sequence design relies on analytic reasoning to restrict attention to a small, tractable subspace. Here, these analytic filtering rules shift from imposing hard design constraints to providing physically motivated performance indicators that guide computational discovery.
\textbf{(d)} A simulation-guided stochastic tree search explores the sequence space through parallel optimization runs. The analytically derived performance indicators provide predictive surrogate evaluation, substantially accelerating the search.
\textbf{(e)} Analytically optimized sequences suppress the dominant decoherence mechanisms, while computational discovery enables further approaching toward the spin-locking limit.
\textbf{(f)} Combinatorially large, discrete sequence spaces. Our approach progressively expands the control alphabet from a small set of symmetry-restricted pulses to include non-Clifford $\pi/3$ rotations, and ultimately to 26,400 hardware-resolved pulse options, as allowed by the control electronics.
}
\label{fig1}
\end{figure*}

Historically, it has been proven remarkably effective to focus attention on analytically tractable subspaces within high-dimensional design landscapes, capture dominant effects, and thus yield constructions approaching performance limits~\cite{Hahn1950,Vandersypen2005,Aliferis2006,Terhal2015}. Outside these subspaces, high-performing solutions are sparse, and performance is highly sensitive to small design changes, which often causes conventional gradient-based algorithms to become trapped at suboptimal points~\cite{qian1999momentum,andrychowicz2016learning}. In near-limit regimes, however, the tractability-imposed boundaries become overly restrictive, by construction excluding unconventional solutions that may better capture weak, performance-limiting effects.
Analytic tractability thus shifts from a guiding principle to a burden.

Here, we address this challenge in pulse-sequence design for many-body spin control, a paradigmatic discrete and combinatorial optimization problem long constrained by analytic tractability.
Since Hahn’s seminal work in 1950, pulse-sequence design has relied on average Hamiltonian theory (AHT) as its analytic foundation~\cite{Hahn1950,Vandersypen2005}, underpinning progress from NMR and MRI to a workhorse control technique across quantum platforms~\cite{Choi2020PRX,Place2021,Burkard2023}.
Although its limitations have long been recognized~\cite{Haeberlen1968,Maricq1982}, pulse-sequence design has remained confined to the tractable subspaces defined by AHT-based structural constraints and screening rules~(Fig.~\ref{fig1}c). This reflects the same circularity: search strategies are formulated within theoretical frameworks, thereby reinforcing confinement to analytically tractable subspaces.

To break this circularity, we repurpose assumptions required for analytic tractability from hard design constraints into soft guidance, and combine them with a stochastic tree search designed for general complex systems~\cite{Wei2025}. Using this physics-guided computational framework~(Fig.~\ref{fig1}d),
we identify a broad class of high-performing pulse sequences that consistently depart from long-standing analytic design principles~\cite{Vandersypen2005,Choi2020PRX}. Experiments in a solid-state spin ensemble confirm substantial coherence enhancements over analytically optimized baselines~(Fig.~\ref{fig1}a,e).
Importantly, these unconventional sequences uncover interpretable structural features that can reliably predict performance, enabling rapid neural network--based evaluation.
The efficiency gain allows expansion of the control alphabet from 8 symmetry-restricted pulses to 26,400 hardware-resolved options, despite the combinatorial scaling of the sequence space~(Fig.~\ref{fig1}f). Within such expanded spaces, fine-grained sequence variation provides the control resolution needed to resolve the impact of weak, hardware-specific effects in a controlled manner. 

These results establish a qualitative shift beyond decades of traditional sequence-design approaches, and highlight a shared
problem structure: near performance limits in realistic quantum hardware, control performance can be constrained by assumptions introduced to preserve analytic tractability.

\section{Analytic Performance Limits}

We study a dense ensemble of approximately $2\times10^{5}$ interacting spins in diamond (see Appendix; Fig.~\ref{fig1}a). Strong disorder, dipolar couplings, and control imperfections give rise to intrinsically complex many-body dynamics in the presence of environmental decoherence (Fig.~\ref{fig1}b).
The control objective is coherence preservation, quantified by the lifetime of prepared spin states under these conditions (see Appendix).

To establish the analytic performance frontier, we benchmark a hierarchy of analytically constructed baseline sequences composed of discrete $\pi$ and $\pi/2$ pulses. Ramsey measurements yield an unprotected decay rate of approximately 500~kHz. The XY family suppresses this to about 40~kHz, while the analytically optimized DROID family further reduces it to roughly 10~kHz~\cite{Vandersypen2005,Choi2020PRX,Gupta2023,Zhou2023PRL}. Continuous spin locking achieves decay rates approaching 1~kHz, setting a practical thermal floor (see Appendix; Fig.~\ref{fig1}e). 

Together, these benchmarks define a clear gap between analytically optimized pulse sequences and the experimentally observed coherence floor set by continuous spin locking. Within reduced, analytically tractable models, the same baselines would be expected to approach this floor. This gap indicates a regime in which tractability assumptions fail to capture residual, performance-limiting effects, such that further improvements can no longer be reliably obtained by extending existing analytic constructions.

To quantify the origins of this gap and establish a reference for subsequent optimization, we construct a coarse-grained decoherence budget and a calibrated simulator. 
The decoherence budget (Fig.~\ref{fig1}e) assumes independent and additive decoherence contributions, following standard practice~\cite{Barry2020,He2023prethermal,Zhang2024blueprint}. It shows that approximately 98\% of the total decay arises from disorder and dipolar interactions, identifying the dominant channels targeted by analytic design principles (see Appendix; Fig.~\ref{fig1}c). In parallel, the calibrated simulator reproduces baseline performance trends with high fidelity (Pearson correlation 0.93, Spearman correlation 0.95; Fig.~\ref{fig2}a; see Appendix). 

\section{Near-Limit Stringent Benchmark}

The analytically optimized DROID family is designed to cancel all leading-order contributions from disorder, dipolar interactions, and pulse imperfections under the framework of AHT~\cite{Choi2020PRX,Gupta2023,Zhou2023PRL}. These sequences do achieve strong experimental suppression of decoherence by mitigating the dominant decay channels. 
The residual 10~kHz decoherence is governed by effects that are either omitted or substantially simplified, and therefore lie beyond the descriptive capability of AHT and reduced models.

This residual-limited regime is precisely the regime where analytic tractability becomes a burden rather than a guide. Systematic guidance for further performance improvement becomes increasingly challenging within closed-form analytic constructions.
Nevertheless, analytic reasoning remains essential: the decoherence budget (Fig.~\ref{fig1}e) and the identification of this residual-limited regime are themselves derived from AHT-based analysis, which establishes the relevant performance scale and frames physical intuition about the dominant mechanisms.
As a result, coherence preservation in interacting spin systems offers a stringent and well-defined benchmark against which sequence-design methods can be evaluated.

\begin{figure*}[tbp]
\includegraphics[width=0.8\textwidth]{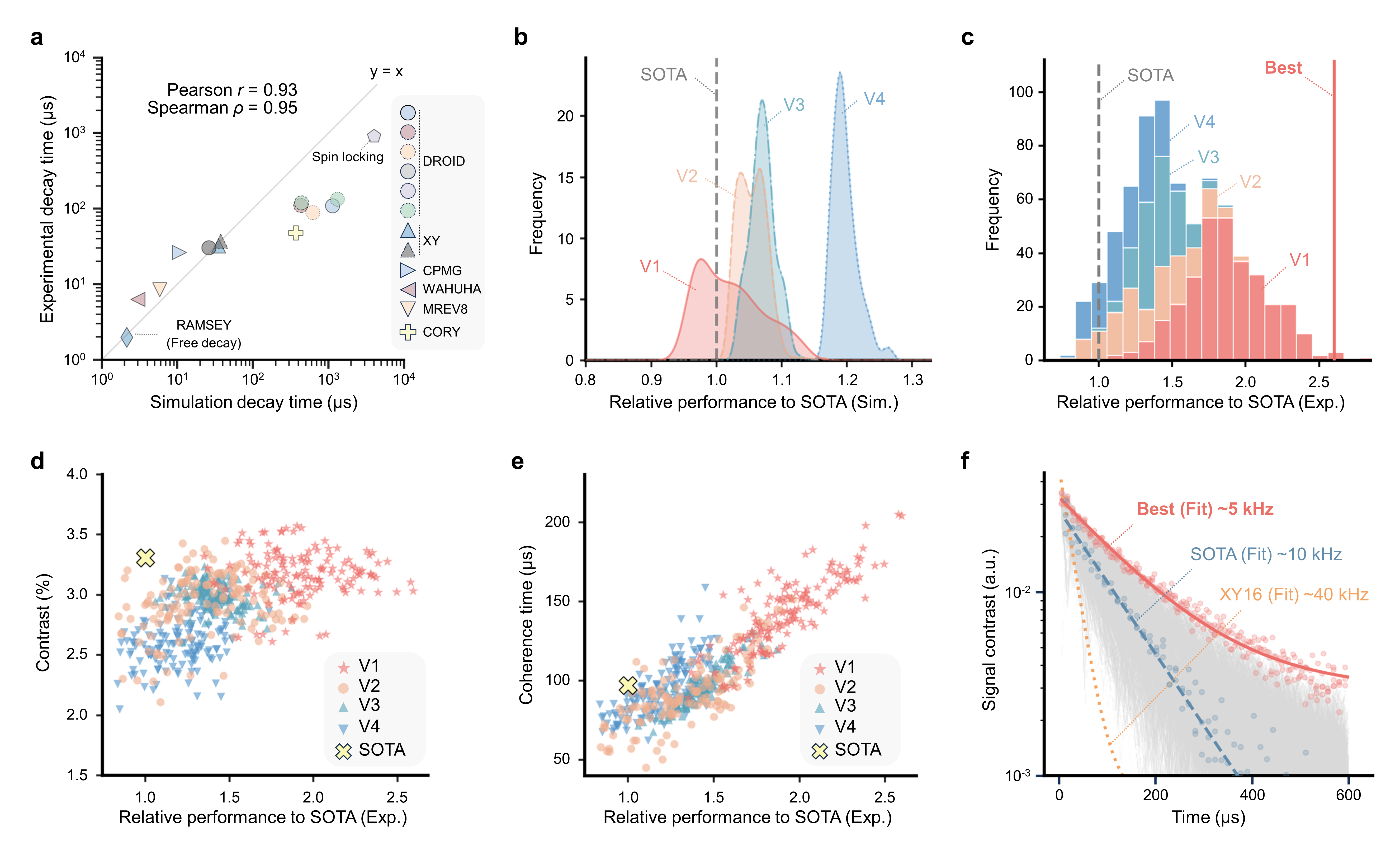}
\caption{
\textbf{Beyond the analytic performance frontier.}
\textbf{(a)} Simulator calibration using Ramsey, XY, DROID, and spin-locking measurements, showing strong agreement between simulated and experimental decay times for baseline sequences (Pearson and Spearman correlation coefficients of 0.93 and 0.95, respectively).
\textbf{(b)} Simulated coherence decay curves for 931 DOESS-discovered sequences across four simulation settings. DOESS identifies sequences that outperform the analytically optimized DROID baseline, with relative improvements of up to $\sim$30\% in Simulator~V4.
\textbf{(c)} Experimental gains are significantly larger, with the highest relative improvement reaching $\sim$150\%.
Sequences are grouped according to the simulator in which they were identified.
\textbf{(d,e)} DOESS sequences exhibit signal contrast comparable to analytic baselines \textbf{(d)} while achieving substantially extended coherence times \textbf{(e)}.
Signal contrast and coherence times are extracted using single-exponential fits.
\textbf{(f)} Experimental coherence decay curves for 931 DOESS-discovered sequences.
The best-performing sequence achieves a decay rate of $\sim$5~kHz, compared to analytically optimized DROID ($\sim$10~kHz) and XY ($\sim$40~kHz) sequences.
}
\label{fig2}
\end{figure*}

\section{Beyond Analytic Performance Limits}

To further approach the spin-locking thermal limit, we use our computational framework, termed DOESS (Data-driven stOchastic tree search that Explores the Sequence Space), to discover control sequences unconstrained by analytic design criteria. At this stage, systematic improvement is limited not by recognizing the constraints imposed by analytic tractability, but by the absence of efficient methods to relax the underlying assumptions.

Within DOESS, AHT-based criteria are repurposed as coarse, physically motivated filters, replacing the overly restrictive binary selection rules of traditional analytic design. Quantitative performance is then evaluated through simulation, free from the assumptions required for closed-form analysis (see Appendix).
To mitigate model–experiment mismatch and improve robustness, multiple optimization rounds are performed in parallel, each guided by a distinct and plausible simulator (Fig.~\ref{fig1}d).

The integration of rapid rejection with robust ensemble evaluation enables our framework to efficiently navigate beyond analytically tractable subspaces defined by AHT (see Appendix).
We define a combinatorial search space of length-24 sequences composed of 13 discrete rotations ($\pi$, $\pi/2$, and $\pi/3$ about the $X$ or $Y$ axes, as well as a no-pulse operation), and relax net-rotation constraints typically imposed for analytic convenience. 
These choices deliberately expand the search beyond symmetry-restricted constructions favored by analytic tractability.
The inclusion of non-Clifford $\pi/3$ pulses generates substantially more complex spin trajectories than Clifford-only constructions (Fig.~\ref{fig1}f). 

Within this largely expanded design space, DOESS consistently identifies sequences that outperform DROID across multiple simulation conditions, with the best candidates exhibiting relative coherence enhancements of up to $\sim$30\% (Fig.~\ref{fig2}b). These improvements are consistent with the trade-off inherent to closed-form AHT analysis, in which higher-order effects within reduced models are neglected.

Experimentally, substantially larger relative improvements are observed, which is up to 150\% (Fig.~\ref{fig2}c), reflecting a pronounced mismatch between reduced models and experimental reality. Such experimental improvement arises from comparable signal contrast combined with a substantially extended coherence time (Fig.~\ref{fig2}d,e). The corresponding raw data are shown in Fig.~\ref{fig2}f, where the longest coherence times increase from approximately 100~\si{\micro\second} to over 200~\si{\micro\second}, or a reduction in the decay rate from $\sim$10~kHz to $\sim$5~kHz (see Appendix).

Of particular note, experimental conditions that are effectively indistinguishable within analytic descriptions give rise to markedly different optimal pulse sequences (see Supplementary Information). This condition-dependent optimality underscores the sensitivity of sequence performance to weak, hardware-specific effects, motivating the development of experiment-in-the-loop optimization approaches capable of learning directly from experimental feedback.

\begin{figure*}[tbp]
\includegraphics[width=0.8\textwidth]{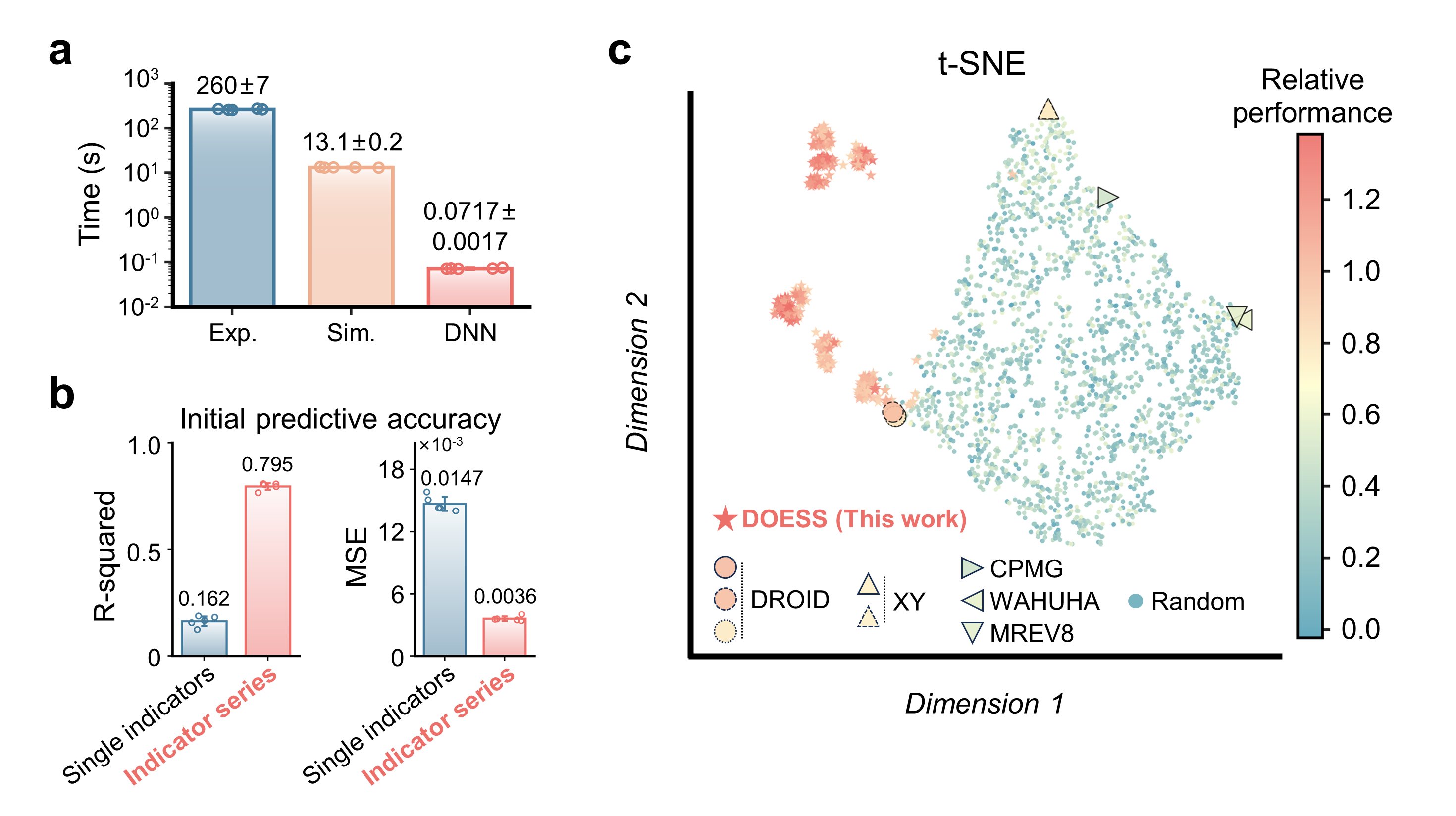}
\caption{
\textbf{Unconventional sequences structures}.
\textbf{(a)} Time required to evaluate a single sequence using experiment (Exp.), simulation (Sim.), and a deep neural network (DNN).
The DNN uses AHT-derived indicator series as input features, with evaluation time dominated by indicator computation.
\textbf{(b)} Comparison of surrogate-model feature representations.
Single-cycle AHT-based performance indicators yield poor predictive performance (best $R^2 = 0.162$, MSE $= 0.0147$), whereas repetition-defined indicator series substantially improve prediction accuracy ($R^2 = 0.795$, MSE $= 0.0036$).
\textbf{(c)} Using indicator series as input features, t-SNE visualization clearly separates DOESS (this work), baseline, and random sequences.
}
\label{fig3}
\end{figure*}

\begin{figure*}[tbp]
\includegraphics[width=0.8\textwidth]{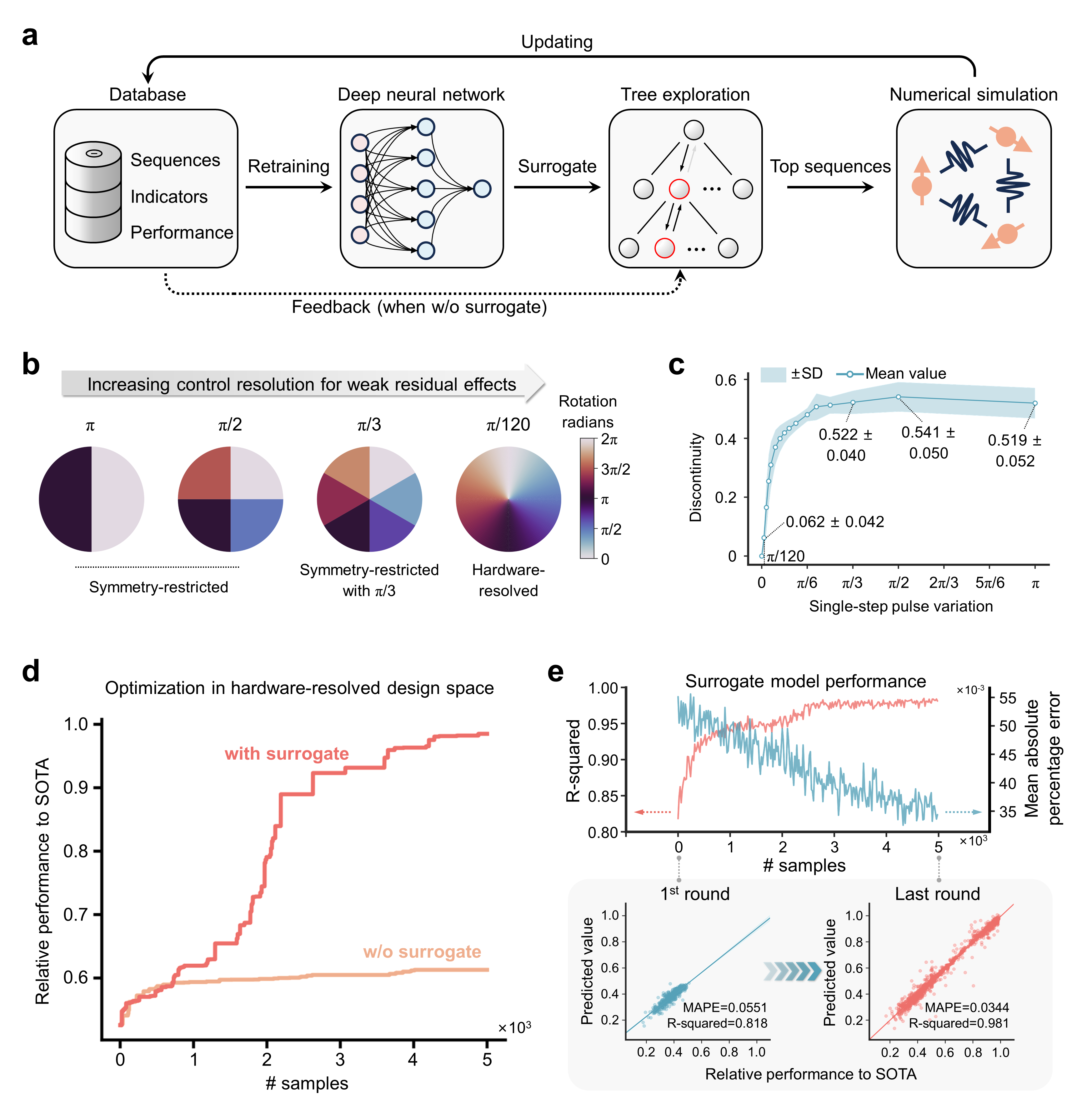}
\caption{\textbf{Predictive surrogate and hardware-resolved sequence space.}
\textbf{(a)} Schematic of the surrogate-assisted optimization loop. In the absence of the surrogate, simulation results are directly fed back to guide the expansion of leaf nodes in the search tree.
\textbf{(b)} Schematic illustrating progressively finer discretization of pulse rotation angles and axes, increasing control resolution needed for addressing weak residual effects.
\textbf{(c)} Finer pulse discretization yields smoother performance variations. Discontinuity is defined as the relative change in simulated performance induced by a rotation applied to a randomly selected pulse. Data are mean $\pm$ s.d. over $n = 5$ instances.
\textbf{(d)} An indicator-enabled predictive surrogate substantially improves search efficiency in the hardware-resolved design space.
For sequence length $n=48$ corresponding to a combinatorial space of over $\sim10^{200}$ possibilities, our approach identifies sequences with performance comparable to the state of the art after evaluating only 5{,}000 candidates.
Without the surrogate, the search stagnates in low-performing regions under the same computational budget.
\textbf{(e)} The surrogate adaptively learns from newly identified sequences during optimization, improving its predictive accuracy from $R^2=0.818$ and mean absolute percentage error (MAPE) $=0.0551$ to $R^2=0.981$ and MAPE $=0.0344$. 
}
\label{fig4}
\end{figure*}

\section{Hidden Structure Beyond Analytic Intuition}

Beyond performance gains, our results reveal a structural mismatch between long-standing analytic intuition and actual near-limit control performance.
Sequences deemed poorly optimized under standard single-cycle diagnostics can nevertheless perform exceptionally well under periodic repetition.
This discrepancy indicates that the limitation lies not in the control protocols themselves, but in the diagnostic scale imposed by analytically convenient single-cycle reasoning.
Specifically, DOESS-discovered sequences systematically deviate from the exact leading-order cancellation conditions enforced by  AHT-based design~\cite{Vandersypen2005,Choi2020PRX}.
These unconventional yet high-performing sequences illustrate how analytic tractability can shift from an effective guiding principle to a limiting constraint.

Under standard AHT-based intuition,
the large single-cycle disorder residuals observed here would be expected to dominate decoherence (see Appendix).
According to the AHT-based decoherence budget, disorder dominates the decay rate (approximately 92\%; Fig.~\ref{fig1}e), and single-cycle AHT indicators heuristically suggest residual decay rates on the order of several tens of kilohertz (see Appendix).
Experimentally, however, the majority of these sequences exhibit decay rates around 10~kHz (Fig.~\ref{fig2}f), substantially lower than this estimate.

Further analysis shows that this apparent mismatch arises because performance is governed by multi-cycle, rather than single-cycle dynamics.
Relaxing net-rotation constraints enables non-identity sequences, for which disorder contributions are determined by the periodically repeated cycle rather than by the single-cycle evolution.
For identity cycles, single-cycle and multi-cycle AHT analyses coincide. 
By contrast, for non-identity sequences accessed by DOESS, repetition defines effective cycles whose time-averaged disorder indicators converge upon repetition and approach values consistent with AHT predictions (see Appendix).

The large residuals observed at the single-cycle level therefore reflect a scale mismatch between local analytic diagnostics and the effective dynamics of the repeated cycle.
For analytic convenience, prior designs predominantly favored identity-cycle 
constructions, reinforcing AHT-based intuition that treats single-cycle 
diagnostics as relevant performance indicators.
The existence of high-performing sequences that tolerate large single-cycle deviations without the anticipated performance penalties reveals a blind spot rooted in the analytic tractability requirements that shape conventional design intuition.

The physical relevance of multi-cycle analysis is further validated through predictive surrogate model training. When trained using single-cycle AHT indicators as input features, the surrogate achieves poor predictive accuracy ($R^2 \approx 0.162$), effectively no better than random guessing. 
By contrast, when trained using indicator series that are computed across repeated cycles, the surrogate attains 
$R^2 = 0.795$ on randomly generated sequences (Fig.~\ref{fig3}b). In addition, using indicator series as input features, t-SNE visualization~\cite{maaten2008visualizing} clearly separates computationally discovered sequences, analytic baselines, and random sequences (Fig.~\ref{fig3}c).
These physics-guided computational representations encode weak residual effects in a predictive and structured manner that departs from conventional closed-form descriptions.

\section{Control Design at Hardware Resolution}

The newly acquired computational capability enables control design at hardware resolution, where increased design freedom allows control protocols to better accommodate the full complexity of realistic hardware.
At the hardware-resolved level, each control pulse is specified by its rotation axis and angle, both discretized in steps of $\pi/120$, yielding over 26,000 distinct pulse operations (Fig.~\ref{fig4}b; see Appendix).

This combinatorial explosion was initially intractable, and the discovery stage was therefore restricted to shorter sequences with a reduced control alphabet consisting of Clifford operations augmented by four $\pi/3$ rotations.
While access to pulse variations at hardware resolution introduces fine-grained control that mitigates single-pulse sensitivity and yields smoother performance landscapes under hardware-specific noise, the fundamental challenge of a highly rugged design space persists.  The corresponding combinatorial growth renders direct application of DOESS at hardware resolution computationally infeasible (Fig.~\ref{fig4}d). More critically, this high-dimensional space remains non-convex and densely populated with local optimum traps. Consequently, gradient-based optimization algorithms are prone to becoming trapped in these suboptimal points, failing to navigate the complex performance topography towards the global optimum.

The predictive surrogate model, enabled by the multi-cycle AHT performance-indicator series, provides sufficiently accurate performance estimates to guide exploration of the vastly larger hardware-resolved space (Fig.~\ref{fig4}a), while suppressing sensitivity to single-pulse perturbations (Fig.~\ref{fig4}c). During optimization, the surrogate model adaptively incorporates newly identified sequences, continuously improving its predictive accuracy (Fig.~\ref{fig4}e).
Using this surrogate-augmented approach, DOESS rapidly identifies candidate sequences in the vastly expanded design space, achieving coherence protection comparable to analytically optimized baselines in simulation (Fig.~\ref{fig4}d).
The resulting pulse sequences generate significantly more intricate spin-evolution trajectories (Fig.~\ref{fig1}f) beyond the reach of closed-form analytic frameworks.

By enabling efficient navigation within hardware-resolved control spaces, this approach establishes a computational foundation for future experiment-in-the-loop optimization, where control strategies can be learned directly from experimental feedback (Fig.~\ref{fig4}a).
The framework is designed to accommodate experimental input by replacing simulation with experiment, but doing so will require further algorithmic innovation in sample-efficient learning to address the substantially lower experimental throughput (Fig.~\ref{fig3}a).

\section{Discussion}

In this work, we identify pulse sequences that systematically depart from long-standing design principles using physics-guided computational discovery. Although deemed poorly optimized under traditional analytic design criteria, these sequences consistently outperform analytically optimized baselines experimentally. Their departure from  conventional design exposes a blind spot arising from a longstanding preference for analytically convenient sequence structures. These computational discoveries thus highlight the discrepancy between analytic tractability and experimental reality.

Beyond experimental performance gains, this work reveals a self-reinforcing structural mechanism underlying decades of traditional pulse-sequence design: assumptions introduced to ensure analytic tractability restrict the accessible solution space, and the resulting restricted solutions, in turn, shape the physical intuition used to justify those same assumptions.
Methodologically, this implies a shift in the role of analytic reasoning, from imposing hard design constraints to providing soft guidance that organizes and informs computational exploration.
Beyond spin control, this problem structure is likely to recur across quantum technologies, wherever simplifying assumptions fail to capture  weak but consequential residual effects near hardware-imposed performance limits.

Looking ahead, optimization within the hardware-resolved sequence space enables the fine-grained control resolution required for precisely targeting weak residual effects, while demonstrating the algorithmic scalability and efficiency necessary for closed-loop experimental optimization. Ultimately, control strategies learned directly from experimental input will be needed to address inevitable model-experiment mismatch, with ensemble-based simulation serving as a practical intermediate route.
Progress toward genuinely experiment-driven optimization will require innovations that tightly integrate algorithmic development with deeper physical guidance, in order to achieve sample-efficient learning under limited experimental throughput. In this context, the physics-guided computational framework demonstrated here offers a promising foundation for further development.

Taken together, these results show that analytic tractability should not be treated as a fixed design principle, particularly when approaching performance limits in realistic quantum hardware. From this perspective, the difficulties of approaching limits arise not merely from intrinsic experimental complexity, but from optimization confined to tractable solution spaces defined by the simplifying assumptions. Physics-guided computational discovery therefore provides a practical pathway for quantum control optimization near hardware-imposed limits, enabling qualitatively different capabilities beyond what closed-form approaches alone can reliably access.

\section*{\label{sec:ack}Acknowledgments}
The authors thank Haixin Qiu and Di Xu for valuable feedback that helped improve the manuscript. Y.Wei and B.P. acknowledge support from CityUHK start-up grant (No.9382006). This work is supported by the EU Horizon Project AMADEUS (101080136); SPINUS (101135699); EU Quantum Flagship Project C-QuENS (101135359); German Federal Ministry of Education and Research (BMBF) through DiaqNOS (13N16462), NeuroQ (13N16487), QSOLID (13N16159), QECHQS  (16KIS1590K), and through the Future Cluster QSens QHMI2 (03ZU2110FB); and Deutsche Forschungsgemeinschaft (DFG) through projects GRK 2642, FOR 2724 (No. 384846402).\\

Y. Wang, Y. Wei, and J. Z. designed the project; R. P., J. Z., and Y. Wang conceived the initial idea; Y. Wei and J. W. managed the project; J. W. provided strategic guidance; Y. Wang developed the numerical simulator; Y. Wei and Y. Wang proposed the ensemble-simulation strategy; Y. Wei and B. P. conducted sequence optimizations and machine learning model training with contribution from H. P.; J. Z. and C. K. C. designed and built the experimental setup, conducted the experiments, and collected all data; J. Z. and Y. Wang analyzed the sequence structures; Y. Wang proposed the idea of physics-informed predictive modeling; B. P. generated the final figures with contributions from J. Z., and G. B.; Y. Wang, J. Z., and B. P. wrote the Supplementary Information; Y. Wang and Y. Wei prepared the initial draft with input from all authors; Y. Wang led the manuscript writing, including rewriting and integration of all components; Y. Wang and Y. Wei supervised the project.

\section*{\label{sec:data}Data and code availability} 
\noindent
All data and code are available as a GitHub repository: \href{github.com/Bop2000/DOESS}
{https://github.com/Bop2000/DOESS}

\appendix

\section{Experimental Setup and Theoretical Modeling}
The diamond sample contains a 1~\si{\micro\meter}-thick NV-rich layer with an $\mathrm{NV}^-$
 concentration of approximately 1~ppm and an isotopic purity of 99.999\% $^{12}\mathrm{C}$ (Element Six Inc.). Optical spin initialization and fluorescence readout are performed using a confocal microscope with a focused laser spot of roughly 1~\si{\micro\meter} in diameter. The corresponding detection volume of $1 \times 1 \times 1$ \si{\micro\meter}$^3$ contains approximately $2 \times 10^{5}$ NV centers.  Such many-body spin systems serve as paradigmatic platforms for quantum sensing~\cite{Barry2024,Barry2020,bucher2019quantum,Gao2025Nature} and simulation~\cite{Choi2019,Zu2021,He2023prethermal,Lei2025}.

For simulation efficiency, disorder is treated as quasi-static, and fluctuations around its mean value are neglected. Spin–spin interactions are modeled as pairwise dipolar couplings. Spin control is implemented through sequences of global microwave pulses that generate coherent spin rotations. Pulse imperfections are modeled in two components: (1) deviations in the implemented rotation angle, leading to under- or over-rotation relative to the intended operation; and (2) the finite pulse duration, during which spins continue to evolve under intrinsic disorder and dipolar interactions. Additional processes such as thermal fluctuations are included only in the theoretical model.

\section{Spin Coherence and Score}
Spin coherence is quantified as the arithmetic average of the survival probabilities of spin states initially polarized along the $X$, $Y$, and $Z$ axes, represented by the state vectors \( \ket{x} \), \( \ket{y} \), and \( \ket{z} \), respectively~\cite{Choi2020PRX}. This definition provides a unified metric applicable across different control protocols, including Ramsey and spin-locking measurements. 

The evolution operator corresponding to a single application of a pulse sequence is denoted by $\mathrm{U}_{\mathrm{total}}$ and is treated as stochastic due to disorder and control imperfections. For an initial state $\ket{\sigma}$, the coherence after $M$ repeated applications of the sequence is averaged over $K$ independent realizations:
\begin{equation}
   \text{Coherence}(T, \sigma) = \frac{1}{K} \sum_{i=1}^K \left|\bra{\sigma} \mathrm{U}_{\text{total},i}^M \ket{\sigma}\right|^2,
   \label{eq:coherence(T)}
\end{equation}
where $\sigma \in \{x, y, z\}$ and \( T \) is the total evolution time determined by the cycle number \( M \). 

To obtain compact performance descriptors, the coherence functions from Eq.~\eqref{eq:coherence(T)} are fit to a single-exponential decay model, \(\text{Coherence}(T,\sigma) \approx C_\sigma e^{-\kappa_\sigma T}\). Here, \( C_\sigma \) is the contrast and \( \kappa_\sigma \) is the decay rate along the axis \( \sigma \). Averaging over the three orthogonal axes yields an overall  decay rate \( \kappa \), and the coherence time is defined as $1/\kappa$, following Ref.~\cite{Choi2020PRX}. 

For use in optimization and ranking, we define a single scalar score that incorporates both contrast and decay by evaluating the coherence at $K$ time points $t_i$:
\begin{equation}
    \text{Performance Score} = \frac{1}{K}\sum_{i}^{K} \left[\frac{1}{3} \sum_{\sigma=x,y,z}  C_{\sigma} \exp\big(-\kappa_{\sigma} t_i\big)\right].
    \label{eq:score}
\end{equation}
When applied to experimental or simulated data, this area under the coherence curve (AUC)-like metric can be computed directly from discrete measurements without assuming a specific decay model. It is therefore robust to non-exponential decay behavior and reduces errors associated with conventional curve-fitting procedures.

\section{Decoherence Budget and Calibrated Simulator}

We consider a set of analytically constructed baseline sequences, including XY8/XY16, DROID, and continuous spin locking, to characterize dominant decoherence mechanisms and to calibrate a numerical simulator. XY8/XY16 sequences primarily suppress disorder-induced dephasing, while DROID further mitigates contributions from dipolar spin-spin interactions and exhibits strong robustness to pulse imperfections. In continuous spin-locking experiments, spins are initialized along an axis in the $XY$ plane and driven continuously by a resonant microwave field. Under the secular approximation~\cite{Wang2023PhdThesis}, perturbations that do not commute with the driving field—including disorder, dipolar interactions, and environmental noise—are strongly suppressed, defining a practical coherence limit. Detailed descriptions are provided in the Supplementary Information: Experimental Details.

Using experimental decay rates obtained from these baseline sequences, we construct a coarse-grained decoherence budget (Fig.~\ref{fig2}a). This budget relies on several simplifying assumptions: (i) pulse errors are assumed to be perfectly cancelled by the analytic sequences; (ii) higher-order contributions from disorder and interactions are neglected; (iii) cross-terms between disorder, interactions, and pulse errors are ignored; (iv) additional environment-induced decoherence channels are not captured; and (v) individual decoherence contributions are assumed to be independent and additive, each modeled by a single exponential decay. The resulting budget therefore provides only an approximate partitioning of decoherence mechanisms, sufficient for calibration but not for predictive accuracy, particularly when near the performance frontier.

To parameterize the experiment-calibrated simulator (Fig.~\ref{fig2}c), we assume certain statistical distributions for the disorder potential and spatial NV positions that determine interaction strengths. These parameters are first tuned to reproduce the experimentally observed decay curves for Ramsey and XY8/XY16 sequences. With these parameters fixed, the strength of pulse imperfections, modeled as a zero-mean Gaussian random variable, is adjusted to match the experimentally measured decay under continuous spin locking. All remaining parameters, including the Rabi frequency and the interpulse free-evolution time, are directly specified by the experimental conditions and held fixed during calibration. Additional details of the calibration procedure and baseline sequences are provided in the Supplementary Information: Experimental Details.

\section{AHT-Derived Performance Indicators}
In coherence-preservation tasks, deviations from ideal control can be quantified by the residual evolution operator $\Delta u$, which captures the accumulated effects of disorder and control imperfections. Conventional AHT-based sequence design treats the leading-order contributions to $\Delta u$ as independent and additive. Under this approximation, individual noise sources are isolated by setting all other imperfections to zero and neglecting higher-order terms. Each resulting cancellation condition defines a binary filtering rule targeting a specific leading-order contribution, including disorder and interactions during pulse application and free evolution, as well as pulse rotation errors.

Because DOESS relaxes these rigid constraints, we replace binary filtering rules with continuous performance indicators that enable more flexible search and ranking. As an example, we consider disorder accumulated during finite-duration pulses. The leading-order contribution is evaluated in a simplified setting where the system Hamiltonian contains only static disorder, \( H = Z \). Pulses are assumed to have finite duration but no rotation-angle errors, and interpulse delays are neglected. From the resulting deviation operator, we extract the effective Hamiltonian \( H' \) generated by the sequence.

Rather than enforcing the strict cancellation condition \( H' = 0 \), we use the Frobenius norm \( \|H'\|_F \) as a continuous performance indicator. Across the 931 DOESS-discovered sequences, this disorder-during-pulses indicator has a mean value of 0.104, corresponding to a 7.3\% residual when normalized by the Frobenius norm of the original Hamiltonian \( H = Z \), which is \( \sqrt{2} \). When the indicator is recalculated for the repeated sequence, its distribution narrows and shifts toward zero, reflecting the cumulative dynamical averaging that emerges upon repetition.

Additional AHT-derived performance indicators and their definitions are provided in the Supplementary Information: Traditional Sequence Design.

\section{Expanded pulse alphabet and search-space construction}
To move beyond the restrictions imposed by AHT-based design, we expand the discrete pulse alphabet by introducing four additional operations that implement $\pi/3$ rotations about the $\pm X$ and $\pm Y$ axes. This increases the number of available pulses from 9 to 13 (including the no-pulse operation) and enlarges the space of length-24 sequences by approximately four orders of magnitude, from $9^{24}$ to $13^{24}$.

The inclusion of $\pi/3$ rotations introduces non-Clifford operations, which transform Pauli operators into linear combinations rather than mapping them within the Pauli group. As a result, the compact algebraic rules that underlie conventional AHT-based filtering no longer apply, rendering analytic pruning strategies ineffective~\cite{Choi2020PRX}. This enlarged discrete sequence space represents a practical compromise between expressiveness and computational tractability prior to the availability of a predictive model.

In the solid-state spin ensemble studied here, there are no fundamental constraints preventing the use of arbitrary rotation angles or rotation axes within the $XY$ plane, provided they are compatible with the resolution of the arbitrary waveform generator (AWG).
Once a physics-informed surrogate model is available (Fig.~\ref{fig3}c), we expand the pulse alphabet to include all hardware-resolved operations permitted by the experimental setup.
Rotation angles and axes are discretized in steps of $\pi/120$, corresponding to the 1~ns time resolution of the AWG, with a minimum rotation angle of $\pi/12$.
This discretization yields a total of 26,400 distinct pulses.
Doubling the sequence length to 48 under this resolution results in a total design space exceeding $10^{210}$ possible sequences.

\section{Ensemble simulation}

Simulations are performed in a Monte Carlo framework, with additional approximations introduced to maintain computational feasibility beyond those used in the analytic modeling. Long-range spin--spin interactions are neglected, and a five-spin subsystem is used to approximate the behavior of the full ensemble. Disorder fluctuations are treated as static, and control imperfections are modeled as quasi-static parameters that vary only between independent simulation runs.

To reduce sensitivity to any single, imperfect model realization, DOESS employs ensemble simulation across multiple plausible parameterizations of disorder strength and control imperfections. Individual simulators can systematically bias absolute performance estimates and ranking orderings due to unavoidable modeling approximations. Ensemble simulation mitigates this effect by increasing the likelihood that at least a subset of simulations captures the residual physics relevant for hardware performance, thereby improving the robustness and transferability of selected candidates. See Supplementary Information: Theoretical Modeling for details.

\section{Neural Network-based performance estimate}
Learning-based models are employed at two distinct stages of the search, serving complementary roles in mitigating the extreme sparsity and ruggedness of the control landscape.

In the initial stage, exploration is restricted to a moderately expanded discrete pulse space composed of $\pi$, $\pi/2$, and $\pi/3$ rotations for sequences of length 24. Within this space, no reliable predictor of sequence performance exists. A two-dimensional convolutional neural network (2D-CNN) is therefore used solely as a physics-informed filter, trained to predict AHT-derived performance indicators. This filter discards approximately 70\% of low-potential candidates prior to costly simulation evaluation, suppressing the overwhelming background of sequences that perform worse than free decay. 
This filtering stage does not attempt to predict coherence performance directly. Instead, it enables the discovery of unconventional high-performing sequences whose structures violate analytically tractable design rules. Analysis of these sequences reveals non-intuitive structural signatures that are not captured by conventional single-cycle AHT metrics, motivating the construction of a physics-informed predictive surrogate.

In the second stage, exploration is extended to the full hardware-resolved control space. Individual pulses are discretized at the experimental resolution, yielding 26,400 distinct operations per step and increasing the sequence length to 48. A physics-informed predictive surrogate is employed to estimate sequence performance and guide stochastic exploration. Numerical simulation is then selectively performed only for candidates whose predicted performance exceeds a predefined threshold.
This surrogate, trained on indicator series evaluated over repeated cycles, provides sufficient predictive accuracy to rank candidates and direct search in regions inaccessible to exhaustive evaluation.  Without this combination of filtering and prediction, systematic exploration at this scale would be infeasible due to the extreme sparsity of high-performing solutions.

Additional implementation details of the predictive surrogate are provided in the Supplementary Information: Neural-Surrogate-Guided Tree Exploration.

\section{Data-Driven Search Tree}
To break the circularity between tractability assumptions and search strategies, DOESS repurposes AHT-derived insight as soft guidance within a stochastic tree search framework. From each node, 24 mutated sequence variants are generated and filtered using AHT-based indicators before numerical evaluation (Fig.~\ref{fig1}a). Evaluated candidates are assigned search scores balancing exploitation and exploration:
\begin{equation}
    \text{Search Score} = \text{Simulation Score} + c_0 \cdot \max(\rho) \cdot \sqrt{\dfrac{2\log N}{n+1}},
\end{equation}
where $N$ is the total node visits, $n$ is the visit count for the current node, and $\max(\rho)$ is the best score observed. This upper confidence bound strategy prioritizes less-visited nodes while retaining information from previous evaluations through backpropagation.
This approach is designed for discrete, combinatorial optimization problems with rugged landscapes and computationally expensive evaluations.
The underlying stochastic tree search algorithm builds on a framework previously developed and validated for learning in complex systems under data-scarce conditions~\cite{Wei2025}. 

Crucially, in this work this general-purpose computational framework is grounded in physical insight derived from AHT. Rather than enforcing analytically motivated construction rules as hard constraints, AHT-based reasoning is incorporated as soft guidance that shapes the search process without restricting the accessible sequence space. The resulting integration of tree search, ensemble simulation, and AHT-guided neural network evaluation proves essential for navigating highly rugged, sparse-reward pulse-sequence landscapes. Implementation details and comparison with other state-of-the-art optimization algorithms can be found in Supplementary Information: Neural-Surrogate-Guided Tree Exploration.

\section{Experimental and Simulation Efficiency}
Photon collection efficiency in the confocal microscope was optimized to minimize the measurement time per sequence while maintaining an adequate signal-to-noise ratio. Including AWG sequence loading, nitrogen nuclear spin polarization, and thermal stabilization, each sequence measurement requires approximately 260 seconds. Complete calibration of all 931 sequences therefore takes about three days of continuous, fully automated data acquisition. 

In simulation, sequence evaluation is substantially faster. Within the same three-day period, DOESS can complete one full optimization cycle for a given simulation setting and evaluate approximately 20,000 sequences (Fig.~\ref{fig2}d), which is roughly 20 times the number acquired experimentally. Simulating a single sequence takes about 10 seconds on a standard laptop equipped with an 8-core, 16-thread AMD R7-5800X CPU. Additional speed-ups of numerical evaluation are achievable through parallelization across multiple machines.

\bibliography{main}

\end{document}